\documentclass[fleqn,twoside,twocolumn]{article}%
\usepackage{amsmath}
\usepackage{amsfonts}
\usepackage{amssymb}
\usepackage{graphicx}%
\usepackage{url}
\usepackage[colorlinks=false]{hyperref}
\usepackage{graphicx}% Include figure files
\usepackage{dcolumn}% Align table columns on decimal point
\usepackage{bm}% bold math
\newcommand{\be}{\begin{equation}}
\newcommand{\ee}{\end{equation}}
\newcommand{\bey}{\begin{eqnarray}}
\newcommand{\bi}{\bibitem}

\begin{document}

\title {Modified theory of gravity and the history of cosmic evolution}

\author{B. Modak$^1$, Kaushik Sarkar$^2$ and Abhik Kumar Sanyal$^3$\\
$^{1,2}$Dept. of Physics, University of Kalyani, Nadia, India - 741235
\\
$^3$Dept. of Physics, Jangipur College, Murshidabad, India - 742213}

\maketitle

\begin{abstract}

A continuous transition from early Friedmann-like radiation era through to late time cosmic acceleration passing through a long Friedmann-like matter dominated era followed by a second phase of radiation era has been realized in modified theory of gravity containing a combination of curvature squared term, a linear term, a three-half term and an ideal fluid. Thus the history of cosmic evolution is explained by modified theory of gravity singlehandedly. The second phase of radiation-like era might provide an explanation to the hydrogen and helium reionization at low redshift.
\end{abstract}

\footnotetext{
\noindent Electronic address:\\
\noindent $^1$bijanmodak@yahoo.co.in\\
\noindent $^2$sarkarkaushik.rng@gmail.com\\
\noindent $^3$sanyal\_ ak@yahoo.com\\}

\section{Introduction}
All the presently available cosmological observations, particularly those at high redshift surveys of supernovae \cite{ries1998, perlmutter1999} and anisotropy of CMBR by WMAP data \cite{sper2003} confirm almost equivocally that the present expansion of the universe is accelerated. The most viable explanation of late time cosmic acceleration is to invoke dark energy dominated universe whose exact nature is not known as yet. Modified theory of gravity \cite{odintsov, Capozziello} appears to unify late time cosmic acceleration with early inflation and hence turned out to be one of the most attractive models in recent years, suitable for explaining the presently available cosmological data. However, a continuous evolutionary picture of the universe from high red-shift till date never appeared in the literature. Here, in the present work, we take up a particular form of action corresponding to modified theory of gravity and present a continuous evolution picture of the universe starting from early radiation era to the late time cosmic acceleration, including the intermediate phase of a long decelerated matter dominated era required to give way to structure formation, started since matter-radiation equality at $z = 3145^{+140}_{-139}$ \cite{komat}, $ 3196^{+134}_{-133}$ \cite{larson}.\\

\noindent
Let us present a brief account of the history of cosmological evolution. After early inflationary era followed by reheating, the universe enters the standard hot Big-Bang radiation dominated era. Universe remains opaque due to Thomson's scattering between thermal photons and the baryons - the electrons in particular. As the universe expands it cools and at around $z \sim 3200$ \cite{komat, larson} matter takes over radiation. Matter and radiation are decoupled with the onset of recombination era at $z_{r} \sim 1080$  \cite{recom:1} and the photons free stream forming the CMBR, as we observe today. The universe in the process becomes transparent giving way to the so called dark age without any source of light. At this epoch a first phase change of hydrogen occurs with the formation of neutral atoms in the universe. During the long course of matter dominated era, the instabilities developed with the evolution give rise to structure formation. The dawn of the universe started at around $z \sim 20$, when the first generation stars and quasars start twinkling. Present observation also suggests that the intergalactic medium (IGM) is filled with ionized plasma. This means, a second phase change of the universe must have occurred at low redshift by some mechanism, when the neutral hydrogen and helium have been ionized. The epoch at which this happened is called the epoch of reionization.  After the phase of reionization the universe starts accelerating. \\

\noindent
Modified theory of gravity is a phenomenological generalization of Einstein's gravity including higher order curvature invariant terms, which plays the role of dynamical dark energy and successfully unifies early inflation with late time cosmic acceleration (see \cite{odintsov} for comprehensive reviews). In particular, such fourth order gravity plays a crucial role as the source of dark energy, since such theories were very successful to explain standard cosmological data such as SNe-Ia fits, an acceleration of the Universe \cite{cap:ga2007}-\cite{browiec2007} and rotation curves for galaxies \cite{rot:cap,rot:mar}. It was also suggested that the standard general relativity together with Dark-Matter and Dark-Energy may be distinguished from $R^n$ approaches with gravitational microlensing \cite{sugg}.\\

\noindent
In the very early universe, a renormalizable theory of gravity \cite{stelle} also requires higher order curvature invariant terms like, $R^2$ and $R_{\mu\nu}R^{\mu\nu}$, in addition to a linear term, generated by one-loop quantum gravitational corrections. Likewise at the end, a particular form of $f(R)$ is therefore also necessary in particular, to establish the claims of modified theory of gravity in the late universe. Such an attempt has been made invoking Noether symmetry. In vacuum or with pressureless dust $f(R) \propto R^{\frac{3}{2}}$ has been found invoking Noether symmetry in the Robertson-Walker line element \cite{cap:vak,vakili}. In fact, despite many possible attempts, e.g. taking into account a scalar-tensor theory of gravity in addition and also considering different anisotropic models \cite{aks:2}, attempting Noether gauge symmetry \cite{an:1a,an:1b} and treating Born-Infeld action being coupled to $f(R)$ \cite{an:2}, no other symmetry has been found to exists for $f(R)$ theory of gravity. Therefore $R^\frac{3}{2}$ is in particular a very special form of $f(R)$ and so it is required to explore the cosmological consequence of such term. Nevertheless, despite claims in favour of such a form of $f(R)$ \cite{cap:plb}, it shows an un-physical evolution like $a \propto t^\frac{3}{4}$ in the radiation era, $a$ being the scale factor. The situation has improved when a linear term is added and it was found to evolve like Friedmann solution ($a \propto \sqrt t$) in the radiation era \cite{aks:2}. However, general analytical solution in the matter dominated era for such an action does not exist. Here, we therefore present numerical solution of the field equations corresponding to an action containing a combination of curvature squared term ($R^2$), a linear term ($R$), a three-half term ($R^{3\over 2}$) and taking both radiation and matter (baryonic and non-baryonic) into account. Note that it is not necessary to incorporate $R_{\mu\nu} R^{\mu\nu}$ term due to the fact that $R_{\mu\nu} R^{\mu\nu} - {1\over 3}R^2$ is a total derivative in $4$-dimension. For the action under consideration, deceleration parameter ($q$) versus redshift ($z$) plot clearly shows yet another radiation era $(q =1)$ in the late universe, in addition to the early radiation era followed by a long Friedmann-like matter dominated era $(q \approx 0.5)$. This late time radiation-like evolution might at least be partially responsible for reionization of neutral atoms present in the IGM. Acceleration of the universe follows thereafter. In the process, the complete history of cosmological evolution from radiation dominated era till date, has been successfully demonstrated.\\

\noindent
In the following section, we construct the model of $F(R)$ theory of gravity, write down the field equations and express them in the form suitable for numerical solution. In section 3, we briefly review the presently available cosmological data. In section 4, we proceed to present numerical solutions, which are depicted in the graphs. In section 5, we demonstrate weak energy limit by transforming the action in canonical form, firstly taking into account an additional tensor degree of freedom and then a scalar degree of freedom. In section 6, we present perturbation equation. Section 7 is dedicated to the understanding of the observed late time radiation era. Section 8 concludes our work.

\section{The model:}
$F(R)\propto R^{1+\delta}$ theory of gravity suffered initial setback under synthesis of light elements, shift of the horizon size at matter-radiation equality and perihelion-precession observation of Mercury \cite{weak}. All these data together puts up severe constraint on $\delta$, viz. $0<\delta<7.2\times 10^{-19}$. Further, solar system also puts up a severe constraints on alternative theories of gravity \cite{st1,st2}. Particularly, for an action
\begin{equation}\label{1}A = \int\sqrt{-g} d^4 x R^{n}\end{equation}
the gravitational potential \cite{sugg} in the weak field limit is expressed as \cite{weak}
\begin{equation}\label{2}\Phi(r) = -\frac{Gm}{2r}\left[1+\left(\frac{r}{r_c}\right)^\beta\right]\end{equation}
where, $r_c$ is an arbitrary parameter varying within the range $(1 - 10^4)$AU, taking into account the velocity of the earth to be $30 ~\mathrm{Km~ s^{-1}}$ \cite{st1} while $\beta$ is related to $n$ as
\begin{equation}\label{3}\beta = \frac{12n^2-7n-1-\sqrt{36n^4+12n^3-83n^2+50n+1}}{6n^2-4n+2}.\end{equation}
Clearly, for $n = 1$, $\beta = 0$, and Newtonian gravitational field is recovered. Any other value of $n$, which appreciably differs from $1$ is ruled out from light bending data in the sun limb and planetary periods \cite{st1}.
The problem was alleviated \cite{odin} by considering an action in the form $\int [\beta R^m + \alpha R + \gamma R^{-n}]\sqrt{-g} d^4 x$, $m>0, n >0$, which passes solar test and therefore is suitable to explain the cosmological evolution right from the inflationary era through to late time accelerated epoch. At the initial stage, $R^m$ term dominates and a de-Sitter solution is realizable for $m = 2$ in particular, explaining inflationary epoch without invoking phase transition \cite{staro,maeda}. In the middle, the linear term dominates giving way to the standard BBN and structure formation and finally $R^{-n}$ term dominates and late stage of accelerated cosmological expansion is realized, without invoking dark energy. However, $R^{-n}$ term is not distinguished at all, since neither it is generated by one-loop quantum gravitational corrections nor from any other physical consequence. Rather, it was considered just to invoke late time accelerated expansion. On the contrary, $R^{3\over 2}$ term appeared as a consequence of Noether symmetry in R-W metric both in vacuum and in matter dominated era \cite{cap:vak}-\cite{an:2}. Further, in contrast to other powers of $R$, no decay of earth radius has been observed for $R^{3\over 2}$ term \cite{brook}. It therefore appears that the gravitational action corresponding to the following form of $f(R)$ theory
\begin{equation}\label{4}\begin{split}&
A =  \int \sqrt{-g} ~d^4 x ~[f(R) + {\cal{L}}_{matter}]\\&
=\int \sqrt{-g} ~d^4 x ~[ \alpha R+ \beta M_{P} R^{3\over 2} + \gamma R^2 + {\cal{L}}_{matter}],
\end{split}\end{equation}
is more suitable to explain cosmic evolution right from the very early stage, till date, since it satisfies all the strong conditions necessary for a viable $f(R)$ theory of gravity. In the above, ${\cal{L}}_{matter}$ is the matter Lagrangian which contains barotropic perfect fluid in the form of radiation and pressureless dust together with CDM. $\alpha (= {M_{P}^2\over 2}=\frac{1}{16\pi G})$, $\beta$, $\gamma$ in the above action stand for dimensionless coupling constants and $\Lambda$ stands for cosmological constant. Here, we would like to mention that Starobinsky model naturally explains inflationary stage and reheating following the mechanism of particle production via scalaron decay, exploiting gravity only \cite{staro, staro1}. In the Starobinsky's action being expressed in Jordan frame
\be S = -\frac{M_p^2}{2}\int\sqrt{-g} d^4 x \left(R - {R^2\over 6\mu^2}\right) + S_m,\ee
where, $\mu = 1.3 \times 10^{-5} M_p$ is a parameter being fixed by the normalization of scalar perturbation amplitude, an additional degree of freedom, viz. scalaron plays the role of inflaton field. The scalaron slow rolls and is responsible for inflationary stage producing a flat power spectrum of perturbation. However, its oscillation reheats the universe. Thus, the action under consideration explains very early stage of cosmological evolution. Here we take up the above action to enunciate the fact that after the reheating is over, the universe being at radiation dominated era, evolves smoothly to a matter dominated era due to the presence of the linear term and a late time acceleration is realized via $R^{3\over 2}$ term. Field equations corresponding to $f(R)$ theory of gravity, viz.
\begin{equation}\label{5}\begin{split} &
(R_{\mu\nu}+g_{\mu\nu}\Box-\nabla_\mu\nabla_\nu)f_{,R}-{1\over 2}g_{\mu\nu}f(R)=T_{\mu\nu}
\end{split}\end{equation}
where $f_{,R}$ is the derivative of $f(R)$ with respect to $R$, now reads for the model (4) under consideration,
\begin{equation}\label{6}\begin{split} &
2 \alpha \left(R_{\mu \nu }-\frac{1}{2}R g_{\mu \nu }\right)\\&+3\beta M_{P}\bigg[\sqrt{R} R_{\mu \nu }
\left.+ \square \sqrt{R} ~ g_{\mu \nu}-\sqrt{R}_{;\mu ;\nu }-\frac{1}{3} R^{3/2} g_{\mu \nu }\right]\\&
+4\gamma \left[ R R_{\mu \nu }+ \square R g_{\mu \nu }- R_{;\mu ;\nu}-\frac{1}{4}R^2g_{\mu\nu }\right]=T_{\mu \nu}.
\end{split}\end{equation}
Note that in the flat Robertson-Walker metric
\begin{equation}\label{7}
ds^2 = -dt^2+a(t)^2\left[dr^2+r^2 (d\theta^2 + \sin^2\theta d\phi^2)\right],
\end{equation}
and in the absence of Einstein-Hilbert term, $R^2$ term and cosmological constant term in the action, the field equation can be expressed in terms of deceleration parameter ($q$) and the Hubble parameter ($H$) as $3\dot{q} + 2(1-q)(1+2q)H =0$. Thus, an analytical solution in the early vacuum dominated universe, when $R^{\frac{3}{2}}$ term dominates over others is given by
\be\label{8}
a(t)^2= \frac{1}{2} \left[ (A t + B )^4 - C^2\right];\;\; q = \frac{C^2- a(t)^2}{C^2+ 2  a(t)^2},
\ee
where $A,~B $ and $C$ are constants. Above solution (\ref{8}) indicates power law inflation and is similar to those presented by Capozziello \cite{cap:plb} and Sarkar et al \cite{aks:1}. Nevertheless, the same solution is admissible even at the late stage of cosmic evolution taking baryonic and dark matter into account \cite{aks:1}. This clearly indicates that $R^\frac{3}{2}$ term is compatible to generate either an inflaton field in the early universe or dark energy in the late. However, $R^{3\over 2}$ term is not an outcome of a renormalizable theory of gravity, rather, as already mentioned, it appears invoking Noether symmetry of $f(R)$ theory of gravity. Therefore it should not be treated to explain inflation. Rather it should be treated as dark energy. Nevertheless, when treated as dark energy, the early radiation and matter dominated era do not track Friedmann like solution, giving rise to the problems in explaining Nucleosynthesis and structure formation. The problem was alleviated by coupling $R^\frac{3}{2}$ with a linear term (Einstein-Hilbert) \cite{aks:2}. The solution obtained in the process \cite{aks:2} tracks Friedmann like evolution in the radiation dominated era. Although exact analytical solution in the matter dominated era was not found, a particular solution indicated late time cosmic acceleration, which is promising. Therefore, in the absence of exact analytical solution of the field equation (\ref{6}), here we simulate numerical solutions taking both radiation and pressureless dust into account.\\

\noindent
Now for the purpose of obtaining numerical solution, we express the field equations in the flat Robertson-Walker metric (\ref{7}) taking Hubble function $H(z)$ as a function of the red-shift parameter $z$. In the process, the trace and the time-time component of the field equations (\ref{6}) (under the choice, $8\pi G = 1, ie., \alpha = {M_{Pl}^2\over 2}={1\over 2}$) are expressed as
\be\label{9}\begin{split} &6H^2\Big[2-{H'\over H}(1+z)\Big]+\frac{3\sqrt 3}{2\sqrt 2}\beta\sqrt {H}\Big[2H-H'(1+z)\Big]^{-\frac{3}{2}}\\&
\Big[64 H^4 +(1+z)^2 \Big(3 (1+z)^2 H'^4 -2 (1+z) H H'^2\big(22 H'\\&
-9 (1+z) H''\big)+3 H^2 \big[47 H'^2+ (1+z)^2\big(2H' H'''\\& -H''^2\big)-18(1+z)H'H''\big]-12  H^3 \left(14(1+z)^{-1} H'\right.\\&
\left.+ (1+z) H'''-4 H''\right)\Big) \Big]-72 \gamma  H (1+z) \Big[(1+z)^2H'^3\\& +H (1+z) H' \Big(4 (1+z) H''-7 H'\Big)+H^2 \Big(6 H'\\&
+(1+z) \big(H''' (1+z)-4 H''\big)\Big)\Big] =(1+z)^3\rho_{m0}
\end{split}\end{equation}
\begin{equation}\label{10}
\Big(\frac{H}{H_0}\Big)^2=\Omega _{\text{m0}}(1+z)^3+\Omega _{\text{r0}} (1+z)^4 + \Omega_{c} ~.
\end{equation}
In the above equations dash ($'$) stands for derivative with respect to the redshift parameter $z$, $\rho_{m0}$ is the present value of matter density, $\Omega _{\text{m0}}$ and $\Omega _{\text{r0}}$ are the present values of matter and radiation density parameters respectively, while $\Omega_{c}$ is the contribution of the higher order curvature invariant terms $R^2$ and $R^{\frac{3}{2}}$ to the density parameter which acts as the source of dynamical dark energy and is given by
\begin{equation}\label{11}\begin{split}&\Omega_{c} =  -\sqrt{\frac{3}{2}}~\beta\left( \frac{H^3}{H_0^2\sqrt{2-\frac{H'}{H} (1+z)}}\right)
\left(3 (1+z)^2 \frac{H''}{H}\right.\\&
\left.+ (1+z)^2 \Big(\frac{H'}{H}\Big)^2-7 \frac{H'}{H}+4\right)-12 \gamma \frac{H^4}{H_0^2}(z+1)\\&
\left((z+1) \left(\frac{2 H''(z)}{H(z)}+\left(\frac{H'(z)}{H(z)}\right)^2\right)-\frac{4 H'(z)}{H(z)}\right)\end{split}
\end{equation}
Additionally, the deceleration parameter $q = -\frac{a\ddot a}{\dot a^2}$ can be expressed in terms of the Hubble parameter or the effective state parameter ($w_e$) as
\begin{equation}\label{12} q =  (1 + z)\frac{H'}{H} - 1 = \frac{3w_e +1}{2}
\end{equation}
which are useful to find numerical solution.

\section{Presently available data:} The present value of the Hubble
parameter is $H_0 = 73.8 \pm 2.5~ Km.s^{-1}Mpc^{-1}$, as reported by Riess et al in 2011 \cite{riess}. In view of its standard form, viz. $H_0 = 100 h~ Km. s^{-1}Mpc^{-1}$, it implies $0.713\le h \le 0.763$. Now under the choice of unit $8\pi G = c = 1$, $H_0 = \frac{h}{9.78}$ $Gyr^{-1}$ and therefore $0.073 \le H_0 \le 0.078$. This means that taking the age of the universe $t_0 = 13.7 Gyr$, $H_0 t_0$ lies within the range, $1 \le H_0 t_0 \le 1.07$, which is fairly good.\\

\noindent
Considering 7-year WMAP data, BAO data and the present value of Hubble parameter $H_0$ altogether, the present value of the effective state parameter has been constrained to $w_{e0} = -1.10 \pm 0.14$ by Komatsu et al in 2010 \cite{komat}. This implies that the present value of the deceleration parameter $q_0$ (obtained in view of equation (\ref{12})) lies within the range $-1.36 \le q_0 \le -1.24$. Note that high redshift type Ia Supernovae has not been taken into account which makes $q_0$ more negative. Further, the range of effective state parameter and correspondingly the deceleration parameter have been fixed in view of $\Lambda$CDM model only, and so these should not be treated as experimental data.\\

\noindent
Before the end of its operation in 2013, the 9-year WMAP \cite{wmap1, wmap2} had measured the third acoustic peak in the temperature power spectrum (TT) with fair precision. As a result, much tighter constraints on the density parameters have been presented by Larson et al \cite{larson}. In the context of the flat $\Lambda$CDM model, the total matter density (which is the sum of the physical baryon density and the cold dark matter density) has been constrained to $\Omega_m h^2 = 0.1334^{+.0056}_{-.0055}$. Thus the density parameter in view of the Hubble parameter data lies within the range $0.2197 \le \Omega_m \le 0.2734$. Allowing tensor modes in the context of $\Lambda$CDM model, the primordial power
spectrum constraints the dark energy density parameter to $0.726 \le \Omega_{\Lambda} \le 0.788$, which is the same as obtained adding axion type iso-curvature perturbation \cite{komat}. Finally, curvaton type iso-curvature perturbation constraints it in the limit $0.738 \le \Omega_{\Lambda} \le 0.794$. All these data together, restricts the matter density parameter to $0.206 \le \Omega_m \le 0.274$. Keeping all these parameters within the specified range, we are now in a position to present numerical solutions of the field equations.
\pagebreak
\section{Numerical solution:}
To obtain $H(z)$ as a solution of the field equation (\ref{9}) containing up-to third derivative, it is required to set three boundary conditions viz. $H_0, H'_0, H''_0$. For this purpose, we undertake the following scheme. Setting the present values of the Hubble and deceleration parameters $H_0$ and $q_0$ by hand, $H'_0$ is obtained in view of equation (\ref{12}). $H''_0$ may then be found in view of equation (\ref{11}), provided the coupling parameters $\beta$, $\gamma$ and the present value of the density parameter $\Omega_{c0}$ are set a-priori.

\subsection {Case - I, [$\beta=2.903$ $\gamma =0.0001$, $H_0=0.074$, $\Omega_{c0} = 0.74$, $q_0 = -0.5$]}
In the present case, we set $\beta = 2.903$, $H_0 = 0.074$ and the limiting present value of the effective state parameter, $w_{e0} = -\frac{2}{3}$, which fixes $q_0 = -0.5$, in view of equation (\ref{12}). Finally the density parameters $\Omega_{m0} = 0.26$ and $\Omega_{r0} = 8 \times 10^{-5}$ are taken into account, which set $\Omega_{c0} = 0.74$. Thus we find $H'_{0} = 0.037$ and $H''_{0} = -0.103458$. Using these parametric values, the trace equation (\ref{9}) is solved numerically and three in one different plots of $q$ versus $z$ (Figure-1) have been presented in both high, medium and low redshift regions.\\

\noindent
Figure-1 depicts that at $z > 3200$ the universe was purely in radiation dominated era ($q = 1,\;w_e = {1\over 3}$) due to the presence of real photons which form CMBR today. From around $z = 3200$, the matter-radiation equality, the deceleration parameter falls off to $q \approx 0.9$ at the decoupling era around $z \approx 1100$ as shown in high redshift plot in figure-1 (right inset). Thereafter, the deceleration parameter $q$ falls sharply with the redshift parameter $z$ to enter exact Friedmann type matter dominated era ($q \approx 0.5,\;w_e = 0$) at around $z = 200$. The deceleration parameter falls a little below $0.4$ and then starts increasing very slowly again at around $z \approx 20$ reaching the peak with $q = 1,\;w_e={1\over 3}$ at around $z \approx 2.5$, as is evident from the low redshift plot (left inset). Transition to an accelerated phase starts around $z \approx 1.39$. Thereafter it crosses the phantom divide line at around $z = 1$ and makes a second transition to come out of it at $z = 0.5$.\\
\pagebreak
\begin{figure}
\begin{center}
\includegraphics[height=6.0cm,width=8.0cm,angle=0]{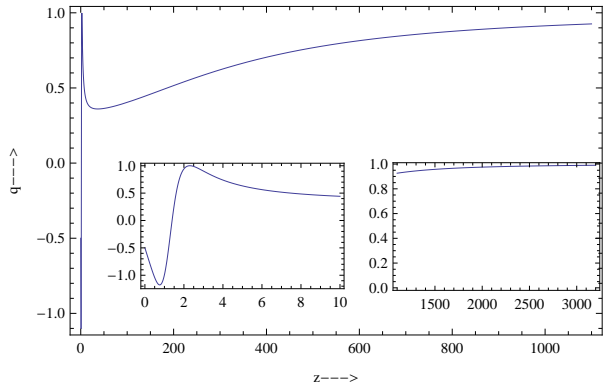}
\caption{The plot $q$ versus $z$ for $\beta = 2.903$ and $\gamma = 0.0001$ (Case-I) depicts that the universe was in pure radiation era at $z > 3200$. Deceleration parameter falls off from the matter-radiation equality epoch to $q \approx 0.9$ at $z \approx 1100$ - the decoupling epoch. It falls even sharply thereafter and a Friedmann type ($q \approx 0.5$) matter dominated era is reached at around $z \approx 200$. The deceleration parameter starts increasing slowly from around $z = 20$ and it is peaked ($q = 1$) at around $z = 2.5$. Late time acceleration starts at around $z = 1.39$. Thereafter it crosses the phantom divide line at around $z = 1$ and makes a second transition out of it at $z = 0.5$.} \label{figure 1}
\end{center}
\end{figure}

\subsection {Case - II, [$\beta=9.3$ $\gamma =0.0001$, $H_0=0.074$, $\Omega_{c0} = 0.74$, $q_0 = -0.6$]}
Here, we increase the value of $\beta$ substantially, so that lower value of the effective state parameter is admissible. To enunciate, we take $\beta = 9.3$, $H_0 = 0.074$ and the present value of effective state parameter, $w_{e0} = -0.733$, which fixes $q_0 = -0.6$, in view of equation (\ref{12}). Finally the density parameters $\Omega_{m0} = 0.26$ and $\Omega_{r0} = 8 \times 10^{-5}$ are taken into account, which set $\Omega_{c0} = 0.74$. Thus we find $H'_{0} = 0.0296$ and $H''_{0} = -0.0609391$. Using these parametric values, the trace equation (\ref{9}) is again solved numerically and three in one different plots of $q$ versus $z$ (Figure-2) have been presented in both high, medium and low redshift regions, as before.\\

\noindent
Figure-2 depicts the same behaviour as figure-1, viz. at $z > 3200$ the universe was purely in radiation dominated era ($q = 1,\;w_e = {1\over 3}$) due to the presence of real photons which form CMBR today. From around $z = 3200$, the matter-radiation equality, the deceleration parameter falls off to $q \approx 0.83$ at the decoupling era around $z \approx 1100$ as shown in high redshift plot in figure-2 (right inset). Thereafter, the deceleration parameter $q$ falls sharply with the redshift parameter $z$ to enter exact Friedmann type matter dominated era ($q \approx 0.5,\;w_e = 0$) at around $z = 250$. The deceleration parameter falls a little below $0.4$ and then starts increasing very slowly again at around $z \approx 20$ reaching the peak with $q = 1,\;w_e={1\over 3}$ at around $z \approx 3.2$, as is evident from the low redshift plot (left inset). Transition to an accelerated phase starts around $z \approx 2$. Thereafter it crosses the phantom divide line at around $z = 1.5$ and makes a second transition to come out of it at around $z = 0.5$. It is to be mentioned that larger value of $\beta$ is required to obtain lower present value of effective state parameter. \\
\begin{figure}
\begin{center}
\includegraphics[height=6.0cm,width=8.0cm,angle=0]{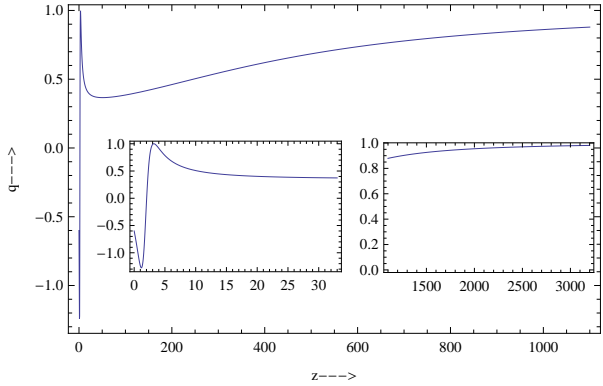}
\caption{The plot $q$ versus $z$ for $\beta = 9.3$ and $\gamma = 0.0001$ (Case-II) depicts that the universe was in pure radiation era at $z > 3200$. Deceleration parameter falls off from the matter-radiation equality epoch to $q \approx 0.83$ at $z \approx 1100$ - the decoupling epoch. It falls even sharply thereafter and a Friedmann type ($q \approx 0.5$) matter dominated era is reached at around $z \approx 250$. The deceleration parameter starts increasing slowly from around $z = 20$ and it is peaked ($q = 1$) at around $z = 3.2$. Late time acceleration starts at around $z = 2$. Thereafter it crosses the phantom divide line at around $z = 1.5$ and makes a second transition out of it at $z = 0.5$.} \label{figure 2}
\end{center}
\end{figure}

\subsection {Case - III, [$\beta=-0.22$ $\gamma =0.000001$, $H_0=0.076$, $\Omega_{c0} = 0.777$, $q_0 = -2.6$]}
Interestingly enough, the same features as above are observed taking even negative values of the coupling parameter $\beta$. For example, choosing $\beta = -0.22$, $H_0 = 0.076$, the present value of the effective state parameter, $w_{e0} = -2.07$ ($q_0 = -2.6$). Finally, taking into account the density parameters $\Omega_{m0} = 0.223$ and $\Omega_{r0} = 8 \times 10^{-5}$, which set $\Omega_{c0} = 0.777$, one finds $H'_{0} = -0.1216$ and $H''_{0} = 1.37486$. Using these parametric values, the trace equation (\ref{9}) is again solved numerically and the plots of $q$ versus $z$ are presented (Figure-3) in both high, medium and low redshift regions.\\

\noindent
Figure-3 as before, depicts that at $z > 3200$ the universe was purely in radiation dominated era ($q = 1,\;w_e = {1\over 3}$) due to the presence of real photons which form CMBR today. From around $z = 3200$, the matter-radiation equality, the deceleration parameter falls off to $q \approx 0.77$ at the decoupling era around $z \approx 1100$ as shown in high redshift plot (right inset). Thereafter, the deceleration parameter $q$ falls sharply with the redshift parameter $z$ to enter exact Friedmann type matter dominated era ($q \approx 0.5,\;w_e = 0$) at around $z = 200$. The deceleration parameter falls a little below $0.5$ and then starts increasing very slowly at around $z \approx 25$ and reaches the peak with $q = 1,\;w_e={1\over 3}$ again at around $z \approx 0.85$, as is evident from the low redshift plot (left inset). Transition to an accelerated phase starts around $z \approx 0.25$.\\
\begin{figure}
\begin{center}
\includegraphics[height=5cm,width=7.4cm,angle=0]{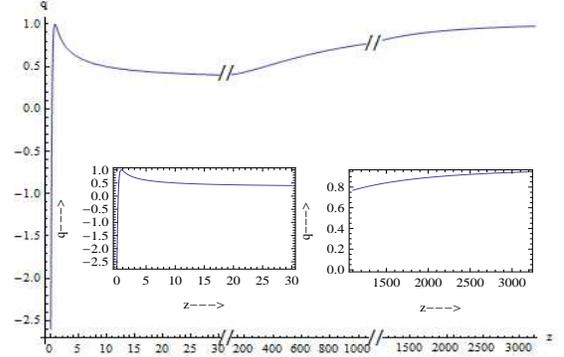}
\caption{The plot $q$ versus $z$ for $\beta = -0.22$ and $\gamma = 0.000001$ (Case-III) depicts that the universe was in pure radiation era at $z > 3200$. Deceleration parameter falls of from the matter-radiation equality epoch to $q \approx 0.77$ at $z \approx 1100$ - the decoupling epoch. It falls off even sharply thereafter and a Friedmann type ($q \approx 0.5$) matter dominated era is reached at around $z \approx 200$. The deceleration parameter starts increasing slowly from around $z = 25$ and it is peaked ($q = 1$) at around $z = 0.85$. Late time acceleration starts at around $z = 0.25$.} \label{figure 3}
\end{center}
\end{figure}

\subsection{Case-IV: [$\beta = -0.22$, $\gamma =0$, $H_0 = 0.076$, $\Omega_{c0} = 0.794$, $q_0 = -2.8$]}
To show that the feature remains unaltered, we have made little change in the data corresponding to case-III, in respect of $\Omega_{c0}$ and $q_0$. With the above data, the boundary conditions $H'_0, H''_0$ have been found as before and the $z$ versus $q$ plot has been presented in figure-4. The figure again depicts that after a long Friedmann-like matter dominated era with $q \approx 0.5$, the deceleration parameter starts increasing and the late time radiation like era $(q = 1)$ is realized at $z \approx 0.8$. Late time acceleration starts at around $z = 0.25$ (inset). Further, matter-radiation equality is clearly visible in the high redshift plot, since the deceleration parameter falls off from its value $q = 1$ at $z = 3200$, to $q \approx 0.79$ at $z = 1100$. \\

\noindent
The feature remains unaltered in the range $-0.15 < \beta < -0.24$ which constraints $0.70 < \Omega_{c0} < 0.81$. Although the chosen present value of deceleration parameter appears to be low, it does not make any problem since as already mentioned, it is model dependent. The feature remains unaltered even for $\gamma < 0$. For example, setting $\gamma=-0.000002$, $\beta = -0.22$, $H_0 = 0.076$ and the present value of the effective state parameter, $w_{e0} = -2.2$ ($q_0 = -2.8$), together with the density parameters $\Omega_{m0} = 0.21$ and $\Omega_{r0} = 8 \times 10^{-5}$, which set $\Omega_{c0} = 0.79$, we find $H'_{0} = -0.1368$ and $H''_{0} = 1.5390$. Except for the fact that the peak $q = 1$, ie. the late time radiation era is realized at around $z \approx 0.94$, the feature remains unaltered. \\
\begin{figure}
\begin{center}
\includegraphics[height=8.0cm,width=7.8cm,angle=0]{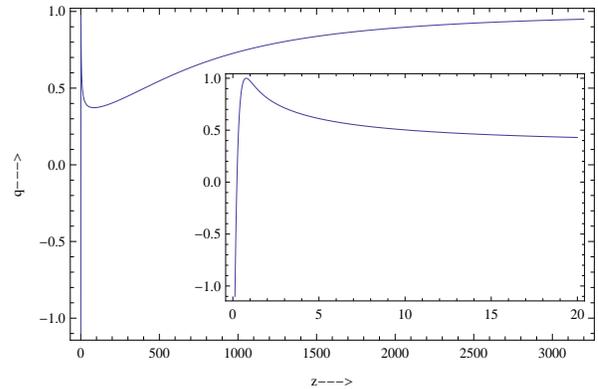}
\caption{The low redshift plot of $q$ versus $z$ for $\beta = -0.22$ (Case-IV) depicts identical feature as in case-III. The only difference is that at low redshift, the deceleration parameter is now peaked $(q = 1)$ at $z \approx 0.8$. Universe then smoothly transits towards late time acceleration starting at around $z = 0.25$ (inset).} \label{figure 4}
\end{center}
\end{figure}

\noindent
Thus, it has been possible to explain the history of cosmic evolution right from the radiation dominated era at $z > 3200$ till date, in the modified theory of gravity containing a linear term, a curvature squared term together with $(R^{\frac{3}{2}})$ term in the presence of an ideal fluid and CDM. A recent Friedmann-type radiation era ($q = 1, w_e = \frac{1}{3}$) is clearly the outcome of the curvature term $R^{\frac{3}{2}}$, since CMBR photons do not play any role at this epoch. \\

\section{Weak energy Limit}
At this stage it is important to discuss the behaviour of $f(R)$ theory of gravity in the weak field limit. $f(R)$ theory of gravity gives rise to fourth derivatives in the field equations. To get rid of such complexity, canonical formulation is necessary under the introduction of an additional degree of freedom. This additional degree of freedom might be a tensor mode obtained under variation of the action with respect to the highest Lie derivative of the extrinsic curvature tensor \cite{Boulware} or the said tensor itself \cite{Horowitz}. It might also be a scalar mode (the scalaron) obtained under scalar-tensor equivalence via conformal transformation \cite{olmo}. Here we shall discuss, the fate of the present model under weak field limit following both the methodology one-by-one.

\subsection{Canonical formulation with a tensor mode}
In a series of articles Sanyal and Modak, Sanyal and his co-workers had developed the formalisms of Boulware \cite{Boulware} and Horowitz \cite{Horowitz} to produce a canonical theory of Einstein-Hilbert action being modified by curvature squared term in Robertson-Walker minisuperspace model \cite{abhik}-\cite{abhik1c}. In particular, canonical formulation of Einstein-Hilbert action
being modified by scalar curvature squared term in Robertson-Walker metric appears in the literature \cite{abhik2} as.
\be A = \int\Big(\dot{h_{ij}}\pi^{ij} + \dot{K_{ij}}\Pi^{ij} - N \mathcal{H}\Big)dt d^3 x,\ee
where, the basic variables $h_{ij}$ and $K_{ij}$ are the metric on 3-space and extrinsic curvature while $\pi^{ij}$ and $\Pi^{ij}$ are canonical momenta respectively. In the above, $N$ is the lapse function, while $\mathcal{H}$ is the Hamiltonian. Here we show that such canonical formulation is also possible for an action containing $R^{3\over2}$ term. For simplicity, we drop out matter and curvature squared term and take up action (\ref{4}), as
\begin{equation}\label{16a}
A_1 = \int \sqrt{-g} ~d^4 x ~[ \alpha R+ \beta_1 R^{3\over 2}]+\sigma_1+\sigma_2,
\end{equation}
where, $\beta_1 = \beta{M_{P}}$ and $\sigma_1 = 2\alpha\int K\sqrt{h}d^3x,~\sigma_2 = 2\beta_1\int K f'(R)\sqrt{h}d^3x$ are the Gibbons-Hawking-York term and the boundary term required to supplement higher order curvature invariant term respectively. Now, under the choice $h_{ij} = a^2 = z$ the Ricci scalar takes the form, $R = {6\over N^2}(\frac{\ddot z}{2z}+N^2\frac{k}{z}-\frac{\dot z \dot N}{2 z N})$ and the above action now reads
\begin{equation}\label{16a1}\begin{split}&
A_1 = \int \left[\frac{\ddot z \sqrt z}{2 N}+N k \sqrt z-\frac{\sqrt z \dot z \dot N}{2 N^2}\right.\\&
 \left.+\frac{\sqrt 3 \beta_1}{2 \alpha N^2}\left(\ddot z-\frac{\dot z \dot N}{N}+2 k N^2\right)^{3\over 2}\right]dt+\sigma_1+\sigma_2.
\end{split}\end{equation}
Under integration by parts the first terms in the above action yields a counter term that gets canceled with $\sigma_1$ and we are left with
 \begin{equation}\label{16b}\begin{split}&
A_1 = \int \left[-\frac{\dot z^2}{4 N \sqrt z}+N k \sqrt z\right.\\&
\left.+\frac{\sqrt 3 \beta_1}{2 \alpha N^2}\left(\ddot z-\frac{\dot z \dot N}{N}+2 k N^2\right)^{3\over 2}\right]dt+\sigma_2.
\end{split}\end{equation}
Now introducing the auxiliary variable as
\begin{equation}\label{16c}
Q = \frac{\partial L}{\partial \ddot z}=\frac{3 \sqrt 3 \beta_1}{4 \alpha N^2}\left(\ddot z-\frac{\dot z \dot N}{N}+2 k N^2\right)^{1\over 2},
\end{equation}
one can express above action in the canonical form as
\begin{equation}\label{16d}\begin{split}&
A_1 = \frac{4}{3}\int \left[Q \ddot z-\frac{\dot N}{N}\dot z Q+2 k N^2 Q-\frac{8 \alpha^2 N^4}{27\beta_1^2}Q^3\right.\\&
\left.-\frac{3 \dot z^2}{16 N \sqrt z}+\frac{3}{4}N k \sqrt z\right] dt+\sigma_2.
\end{split}\end{equation}
Now the first term in (\ref{16d}) is integrated by parts and the total derivative term gets canceled with $\sigma_2$. We are then finally left with
(the overall constant term has been absorbed in the action)
\begin{equation}\label{16e}\begin{split}&
A = \int \left[-\dot Q \dot z-\frac{\dot N}{N}\dot z Q+2 k N^2 Q-\frac{8 \alpha^2 N^4}{27\beta_1^2}Q^3\right.\\&
\left.-\frac{3 \dot z^2}{16 N \sqrt z}+\frac{3}{4}N k \sqrt z\right] dt.
\end{split}\end{equation}
The canonical momenta are,
\begin{equation}\label{16f}
p_z=-\dot Q-\frac{\dot N}{N}Q-\frac{3 \dot z}{8 N \sqrt z},~p_Q=-\dot z,~p_N=-\frac{\dot z}{N} Q,
\end{equation}
and the Hamilton constraint equation is,
\begin{equation}\label{16g}\begin{split}&
H_c = -\dot Q \dot z-\frac{\dot N}{N}\dot z Q-\frac{3 \dot z^2}{16 N \sqrt z}-2 k N^2 Q\\&
+\frac{8 \alpha^2 N^4}{27\beta_1^2}Q^3-\frac{3}{4}N k \sqrt z.
\end{split}\end{equation}
In view of the definitions of canonical momenta (\ref{16f})
\begin{equation}\label{16h}
p_Q p_z=\dot Q\dot z+\frac{\dot N}{N}\dot z Q+\frac{3 \dot z^2}{8 N \sqrt z}
\end{equation}
one obtains the following relation,
\begin{equation}\label{16i}\begin{split}&
-\dot Q\dot z-\frac{\dot N}{N}\dot z Q-\frac{3 \dot z^2}{16 N \sqrt z}=-p_Q p_z+\frac{3 \dot z^2}{16 N \sqrt z}\\&
=-p_Q p_z+\frac{3}{16 N \sqrt z}p_Q^2
\end{split}\end{equation}
which allows to express the Hamiltonian constraint equation in terms of the phase space variables as
\begin{equation}\label{16j}\begin{split}&
H_c = -p_Q p_z+\frac{3}{16 N \sqrt z}p_Q^2-2 k N^2 Q+\frac{8 \alpha^2 N^4}{27\beta_1^2}Q^3\\&
-\frac{3}{4}N k \sqrt z=0.
\end{split}\end{equation}
Now in order to express the Hamiltonian in terms of the basic variables, let us choose
\begin{equation}\label{16k}
x=\frac{\dot z}{N},~~Q=\frac{\partial A}{\partial \ddot z}=\frac{\partial A}{\partial \dot x}\frac{\partial \dot x}{\partial \ddot z}=\frac{p_x}{N}~$and$~p_Q=-\dot z=-N x
\end{equation}
to express equation (\ref{16j}) as
\begin{equation}\label{16l}\begin{split}&
H_c = N\left(x p_z+\frac{3}{16 \sqrt z}x^2-2 k p_x+\frac{8 \alpha^2}{27\beta_1^2}p_x^3-\frac{3}{4}k \sqrt z\right)\\&
 = 0 = N\mathcal{H},
\end{split}\end{equation}
It is now straightforward to express the action (\ref{16d}) as [since $\dot z = Nx$; therefore, we substitute $\ddot z= N\dot x + \dot N x$ in the first term of (\ref{16d}), $\dot z = Nx, Q = {p_x\over N}$ in the second and third terms, $p_x^3= N^3Q^3$ in the fourth and $x = {\dot z\over N}$ in the fifth]
\begin{equation}\label{16m}\begin{split}&
A = \int \left[\dot z p_z+\dot x p_x-N \mathcal{H}\right] dt ~d^3x\\&
 = \int\Big(\dot{h_{ij}}\pi^{ij} + \dot{K_{ij}}\Pi^{ij} - N \mathcal{H}\Big)dt d^3 x,
\end{split}\end{equation}
which is the required canonical form, where in addition to the three-space metric $h_{ij}$, the extrinsic curvature tensor $K_{ij}$ play the vital role towards canonical formulation. Apart from the two familiar mass-less spin-2 gravitons arising out of the linearized field energies of these particle excitations, the additional degree of freedom leads to a pair of massless spin-2 particles. Therefore, the model under consideration does not contain ghost degree of freedom. It is now required to check if action (\ref{4}) admits Newtonian gravity so that it might satisfy solar test under weak field approximation which is valid at low energy limit. For this purpose, one can always set $\gamma = 0$, since in no way $R^2$ term influences the solar test. In weak field approximation $g_{\mu\nu}=\eta_{\mu\nu} +h_{\mu\nu}$, where $\lvert h_{\mu\nu}\rvert \ll1$. Retaining only linear terms in $h_{\mu\nu}$ we have
\be\label{13} R_{\mu\nu}\simeq \frac{1}{2}\Box h_{\mu\nu}~~ \mathrm {and}~~ R \simeq
\frac{1}{2} \Box h,~~ \mathrm {where}~~ h = h^{\mu}_{~~\mu}.\ee
The time-time component of field equation is
\begin{equation}\label{14}\begin{split}&
\big(R_{00}-\frac{1}{2} g_{00}R\big)+\beta_1 R^{1/2}(3R_{00}-R g_{00}) +
\frac{3\beta_1}{2}R^{-\frac{3}{2}}\\&
\Big[\big(R\Box R-\frac{1}{2}R_{;\lambda}R^{;\lambda}\big)g_{00}
- R R_{;0;0}+\frac{1}{2}R_{;0}R_{;0} \Big] =  T_{00}.
\end{split}\end{equation}
In static background spacetime, equation (\ref{14}) with only linear term in $h_{\mu\nu}$ yields (terms containing derivatives of $R$ have been discarded as they will contain third and fourth derivatives of $\Phi$, which will have no counterparts in Poisson equation.)
\begin{equation}\label{15}
\triangledown^2 h_{00}\simeq \rho.
\end{equation}
or considering next higher order term in $h_{\mu\nu}$, equation (\ref{14}) gives
\begin{equation}\label{16}
\triangledown^2 h_{00} + 3\beta_1 ~\sqrt{\frac{1}{2}\triangledown^2 h}~\Big(\triangledown^2 h_{00}-\frac{1}{6}\triangledown^2 h\Big) \simeq \rho,
\end{equation}
(see appendix for detailed calculation). Since at low energy limit Poisson equation is obtained, as in the case of general theory of relativity, so Newtonian gravity is valid at weak energy limit. This is one important technique to test the viability of $f(R)$ theory of gravity under weak energy approximation.\\

\subsection{Canonical formulation with a scalar mode - the Chameleon Mechanism}
Canonical formulation of $f(R)$ theory of gravity is also possible via scalar-tensor equivalence. Usually, such a formally equivalent theory is dealt with, to get information regarding the weak field limit of $f(R)$ theory of gravity. For example, the action
\be A = \int\alpha f(R)\sqrt{-g} d^4 x\ee
may be cast in the following Brans-Dicke form of action without the help of conformal transformation
\be A = \int\sqrt{-g} d^4 x[\phi R -V(\phi)],\ee
where, $V(\phi) = \phi\chi -f(\chi)$ and $\chi = R$. Clearly, one observes that this analogy has been established at the cost of vanishing Brans-Dicke parameter $\omega$. Since it is well-known that Brans-Dicke parameter should be large enough and particulary $\omega\to \infty$, to satisfy solar constraint, so under conformal transformation $f(R)$ theory fails to satisfy solar test. For this reason $f(R)$ theory of gravity had initially been ruled out. However, rigorous calculation of Newtonian limit of $f(R)$ theory of gravity taking into account correct analogy between $f(R)$ and scalar-tensor theory, has proved that it is too early to make final conclusion \cite{stabile} as there are other techniques to establish scalar-tensor equivalence. One such technique is Palatini formalism, in which canonical formulation reduces the field equations to second order by considering metric and connection as independent variables. Although Palatini formalism is identical to the metric formalism for general theory of relativity, it differs by and large for higher order theory of gravity. Particularly, scalar-tensor equivalence has been established with a non-zero Brans-Dicke parameter \cite{mota, amar}. Thus Solar test might not fall short in this formalism. This raised interest to understand the situation deeply under metric variation formalism also, which is our present concern.\\

\noindent
Another way to establish scalar-tensor equivalence is possible under conformal transformation \cite{olmo}, which again replaces higher (fourth) order theory to second order, by the introduction of a scalar degree of freedom, dubbed as scalaron. In this technique, the action (33) under a conformal transformation $g_{\mu\nu} \rightarrow f_{,R}g_{\mu\nu} = e^{-2\eta\phi\over M_{Pl}} g_{\mu\nu}$ reads
\be A = \int\left[\alpha R - {1\over 2}\partial_{,\mu}\phi\partial^{,\mu}\phi - V(\phi)\right]\sqrt{-g}d^4 x\ee
with $\eta = -{1\over\sqrt 6}$ and $V(\phi) = \alpha {(Rf_{,R}-f)\over f_{,R}^2}$.
In the process, a technique dubbed as chameleon mechanism had been invoked. Here our aim is to check if under chameleon mechanism our present model passes solar test. For this purpose, following \cite{hojjati} we express action (4) as
\be\label{def}\begin{split}& A = \int\left[\sqrt{-g}d^4x\big(\alpha R + \beta M_P F(R)\big)\right]\\& F(R) = R^{3\over 2}+{\gamma R^2\over \beta M_P}.\end{split}\ee
and compute the trace of the corresponding field equation as,
\begin{equation}\label{18a}
\Box F_{,R} = \frac{1}{3}\left[2F(R) - RF_{,R} + \frac{\alpha R}{\beta M_P}\right] + \frac{T}{6\beta M_P}.
\end{equation}
Expressing the above equation as $\Box F_{,R} = \frac{\partial V_e}{\partial F_{,R}}$, where, $V_e$ is the effective scalaron potential, the mass of the scalaron field may be calculated as
\begin{equation}\label{18b}
m_{F}^2 = \frac{\partial ^2 V_e}{\partial F_{,R}^2} = \left[\frac{\beta M_P F_{,R} + \alpha}{3M_P\beta F_{,RR}} - \frac{R}{3}\right]\end{equation}
In view of the definition of $F(R)$ given in (\ref {def}), the scalaron mass corresponding to the present model in Jordan frame reads,
\begin{equation}\label{me}
 m_{F}=\left[{R\over 3}+{4\alpha\over 9\beta M_P}\sqrt{R}\right]^{1\over 2}
\end{equation}
The same above expression (\ref{me}) may also be obtained considering the wave equation in the Einstein frame under conformal transformation, following \cite{Gannouji} and then translating it back to Jordan frame, by multiplying $m_{Einstein}^2$ by $f_{,R}$. However, for this purpose, we need to take $F(R) = R + 16\pi G(\beta M_P R^{3\over 2} + \gamma R^2)$, instead. Now to study the viability of the chameleon mechanism we need to compare the masses of the scalaron both on earth and at the bulk (cosmological scale). For this purpose, we need to know the value of the Ricci scalar $R$ on earth and on the bulk. For the sake of simplicity, we take help of the Friedmann equation to estimate $R$, which for pressureless dust $(p=0)$ reads
\be R = {\rho\over M_{P}^2}.\ee
Now in air $\rho = 10^{-3} g/cc$, while in the bulk, it is $10^{-29} g/cc$. In the unit $c = 1$, the value of the Ricci scalar on earth and on the bulk may be calculated in view of equation (40) as $R_e \approx 10^{-40} eV^2$ and $R_b \approx 10^{-66} eV^2$ respectively. Therefore, taking the value $\beta = 2.903$, as in case-I, we find the mass of the scalaron on earth to be $m_F(earth) \approx 1.49\times 10^3 eV$. Corresponding Compton wavelength is $\lambda (earth) \approx 1.32\times 10^{-7} mm$, which is negligibly small to produce any correction to the Newtonian gravity. In contrast, the mass of the scalaron on the bulk is $m_{F} (bulk) \approx 4.72\times10^{-4} eV$, which is seven order of magnitude smaller than that on earth. Corresponding Compton wavelength is $\lambda (bulk) \approx 0.42 mm$. Taking case-II on the contrary, for which $\beta = 9.3$, the mass of the scalaron on earth is found to be $m_F(earth) \approx 8.34\times 10^2 eV$. Corresponding Compton wavelength is $\lambda (earth) \approx 2.36\times 10^{-7} mm$, which is negligibly small to produce any correction to the Newtonian gravity. The mass of the scalaron on the bulk, on the other hand is $m_{F} (bulk) \approx 2.638\times10^{-4} eV$, producing compton wavelength $\lambda (bulk) \approx 0.75 mm$. It is important to mention that quantum stability bound gives $5\times 10^-13 eV$ as the lower limit to the mass of the scalaron on bulk \cite{Gannouji}. Although the Compton wavelength corresponding to the bulk is not appreciably large in either case, but we have observed from the graphs (I through IV) that Friedmann solutions have been modified appreciably indicating possibility for long range interaction. It is important to mention that the scalaron mass obtained considering $R^{-1}$ theory of gravity is of the order of $10^{-34}eV$. This value is too small and the corresponding Compton wavelength is larger than the size of the universe and is ruled out by quantum stability criterion \cite{Gannouji}. On the contrary, the mass of scalaron at bulk in the present model is at par with quantum stability bound \cite{Gannouji}. Although $\beta <0$ shows the same cosmological behaviour and can not be ruled out following weak energy limit studied in section (5.1), however, the scalaron mass is negative and therefore is plagued with tachyon or ghost degree of freedom. Thus, the present model passes the solar test with confidence for $\beta > 0$.

\section{Perturbation about background curvature:}
It is believed that higher order theory of gravity modifies deeply the spectrum of perturbation. Therefore let us study the issue in brief. Taking $R = R_b + R_p$, where, $R_b$ and $R_p$ are the background and perturbed curvature scalars respectively, the dynamics of perturbed curvature scalar has been evaluated by Nojiri and Odintsov \cite{odintsov}. In the present case it reads (in the absence of $R^2$ term i.e. $\gamma = 0$),
\begin{equation}\label{17} \ddot R_p + 3H \dot R_p + R_p V(R_b) = 0,\end{equation}
where,
\begin{equation}\label{18} V(R_b) = \frac{1}{2}\frac{\dot R_b^2}{R_b^2} + \frac{4}{9}R_b - \frac{2}{9\beta_1}\sqrt R_b.\end{equation}
Equation (\ref{17}) implies that perturbed space-time is oscillatory with decaying amplitude, suggesting that the background space-time remains unaffected.

\section{A possible interpretation of late time radiation era}
Since everything is well behaved, it is therefore important to make a thorough study to understand the consequence of gravity including $R^{\frac{3}{2}}$ term on the late time cosmic evolution, particularly the nature of the graphs at low red-shift where the peak ($q = 1$), corresponding to the late time radiation era is found.

\subsection{The field equation}
It is important to note that all the important aspects of higher order gravity have been explored only through scalar-tensor equivalence under conformal transformation or using an auxiliary variable $\chi = R$, as already demonstrated. This reduces the theory to a minimally or non-minimally coupled scalar-tensor theory of gravity and the scalar is treated as a real scalar field. Likewise, here we reduce higher order theory under consideration to linear gravity being non-linearly coupled to a tensor field, viz, the elctromagnetic field exhibiting and establishing the equivalence. It is also important to note that no transformation is necessary for this purpose, rather one can simply cast the field equation (\ref{6}) in the following interesting form
\begin{equation}\label{19}G_{\mu\nu} = \frac{1}{2\alpha + 3\beta_1 \sqrt R +4 \gamma  R}\left[T_{\mu\nu} + \Lambda_e(R)g_{\mu\nu} +\mathbb{{T}}_{\mu\nu}\right]
\end{equation}
where, $G_{\mu\nu}$ is the Einstein's tensor, $T_{\mu\nu}$ is the energy-momentum tensor corresponding to matter Lagrangian. $\Lambda_e(R) = -\frac{\beta_1}{2}\Big(R^{\frac{3}{2}} + \frac{9}{2}\Box{\sqrt R}\Big) -\gamma \left(R+3\Box R\right)$ acts as an effective dynamical cosmological constant, $\Box$ being the D'Alembertian operator. Finally, the last and the most interesting term $\mathbb{{T}}_{\mu\nu}$, given by
\begin{equation}\label{20}\begin{split}&
\mathbb{{T}}_{\mu\nu} = 3 \beta_1 \Big(\sqrt {R}_{;\mu;\nu} -\frac{1}{4} \square{(\sqrt R)} g_{\mu\nu}\Big)\\&
+4\gamma \left(R_{;\mu ;\nu }-\frac{1}{4}g_{\mu \nu }\Box R\right).
\end{split}\end{equation}
is clearly traceless. Further, one can show trivially that, under the choice, $F_{\sigma\delta} F^{\sigma\delta} = \frac{3}{2}\Box \sqrt R$,
\begin{equation}\label{21}
 \mathbb{{T}}_{\mu\nu} = E_{\mu\nu} = F_{\mu}^{\sigma} F_{\nu\sigma} - {1\over 4}F_{\sigma\delta} F^{\sigma\delta}
\end{equation}
The term $ (\sqrt R)_{;\mu;\nu}$ is a symmetric tensor, so $F_{\mu}^{~\sigma} F_{\nu\sigma}$ is also symmetric in view of equation (\ref{21}) which holds for both the symmetric and antisymmetric nature of the tensor $F_{\mu\nu}$. Assuming it to be antisymmetric with the choice $F_{\mu\nu}= A_{\mu ; \nu} -A_{\nu; \mu}$, $F_{\mu\nu}$ and $E_{\mu\nu}$ may be interpreted as the field tensor and the energy-momentum tensor of an electromagnetic field respectively, under appropriate choice of unit. Thus it can be shown that
\begin{equation}\label{22}
 \dot{A}^{\mu}_{~;\mu}=  J_{\mu}A^{\mu}+ \frac{1}{2}\Box(A_{\delta}A^{\delta}) -\frac{3}{4}\Box R
\end{equation}
where, $\dot A^{\mu} = A^{\mu}{_{;\sigma}} A^{\sigma}$. $\mathbb{{T}}_{\mu\nu}$ therefore looks very much like an energy-momentum tensor equivalent to that of a source-free ($J_{\mu} = 0$) or interaction free $(J_{\mu}A^{\mu} = 0)$ electro-magnetic field, satisfying the relation $F^{\sigma\delta}A_{\sigma} = \frac{1}{2}(\sqrt R)^{;\delta}$. Note that since $\sqrt R$ is constant on the space-like surface, so $F^{\sigma\delta}A_{\sigma}$ is a time-like vector. With this understanding, one can now clearly observe that if the effective cosmological constant term $\Lambda_{e}(R)$ dominates at the early epoch, inflation would be realized. In the middle, if the perfect fluid energy momentum tensor $T_{\mu\nu}$ dominates, then due to the presence of the interaction term ($2\alpha + 3\beta_1 \sqrt R + 4\gamma R$), a continuous  transition from Friedmannn-like radiation dominated era ($a \propto \sqrt t$ with the effective state parameter $w_e = \frac{1}{3}$) to matter dominated era ($w_e \rightarrow 0$) would be realized. At the later stage of cosmological evolution, if the effective electro-magnetic energy-momentum tensor $\mathbb{T}_{\mu\nu}$ dominates, then one should expect yet another phase of radiation era ($w_e = \frac{1}{3}$). Finally, at the very late stage of cosmological evolution, if the effective electro-magnetic energy-momentum tensor $\mathbb{T}_{\mu\nu}$ falls off sharply at a much faster rate than the effective cosmological constant $\Lambda_e$, so that $\Lambda_e$ again overtakes $\mathbb{T}_{\mu\nu}$, then an accelerated expansion might be realized. These facts have been demonstrated in the spatially flat Robertson-Walker line element in the figures 1 through 4.

\subsection{Gravitational wave equation: }
In this section, we explore to understand the role of the typical late time radiation era on cosmic evolution. For this purpose, let us construct gravitational wave equation assuming linear term in $h_{\mu\nu}$ in the left hand side of (\ref{19}). Thus under the gauge condition $(h^{\mu}_{~\nu} - \frac{1}{2}h~\delta^{\mu}_{~\nu})_{,\mu} = 0$, (\ref{19}) effectively yields,
\begin{equation}\label{25}\begin{split}&
\frac{1}{2} \Box\left( h_{\mu\nu}-\frac{1}{2}h\eta_{\mu\nu}\right)= -3\beta_1\left[\frac{1}{4}g_{\mu\nu}(\sqrt{R})^{;\lambda}_{~;\lambda}\right.\\&
\left.-(\sqrt{R})_{;\mu;\nu}\right]
-\frac{\beta_1}{2}\left(R^{\frac{3}{2}}+\frac{9}{2}(\sqrt{R})^{;\lambda}_{~;\lambda} \right)  g_{\mu\nu} \\&
-\gamma \left(Rg_{\mu \nu }-4 \left(R_{;\mu ;\nu }-g_{\mu \nu }\Box R\right)\right)+ T_{\mu\nu}  =J_{\mu\nu}
\end{split}\end{equation}
where $g_{\mu\nu}= \eta_{\mu\nu} +h_{\mu\nu} $, $\mid h_{\mu\nu}\mid < 1$. In the wave equation (\ref{25}), $J_{\mu\nu}$ acts as the source term which produces the gravitational wave and its behavior depends on its strength. More precisely, the amplitude of the gravitational wave should be calculated from the transverse-traceless part of the space-space component of the energy-momentum tensor, which is the source of the gravitational waves. Here, in figure-5, we therefore present a plot of space-space component $J^1_{~1}$ versus the redshift parameter $z$ corresponding to the Case-II (neglecting the contribution of energy momentum tensor of ideal fluid). The peak at $z \approx 3.2$ of the source term is of the order of $({J^1_{~1}})^{\frac{1}{4}} \simeq T \approx 60 ~eV$ and are located around the peak of the deceleration parameter $q = 1$. In figure-6, we present a plot of space-space component $J^1_{~1}$ versus the redshift parameter $z$ corresponding to case-IV. The peaks at $z \approx 0.7712$ and $z \approx 0.7722$ of the source term are of the order of $({J^1_{~1}})^{\frac{1}{4}} \simeq T \approx 48 ~eV$ and are located around the peak of the deceleration parameter $q = 1$.
\begin{figure}
\begin{center}
\includegraphics[height=4.0cm,width=6.8cm,angle=0]{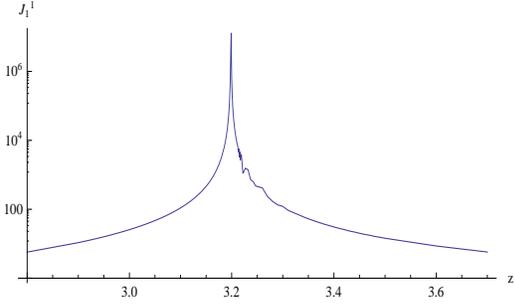}
\caption{A plot of $J^1 _{~1}$ in $eV^4$, versus $z$ (Case-II) shows a peak at $z \approx 3.2$ where, $J^1_{~1} = 1.3 \times 10^7 ~eV^4$, i.e. $(J^1_{~1}) ^{\frac{1}{4}} \simeq T \approx 60 ~eV$, which is sufficient to reionize intergalactic medium.}\label{figure 5}
\end{center}
\end{figure}
\begin{figure}
\begin{center}
\includegraphics[height=4.0cm,width=6.8cm,angle=0]{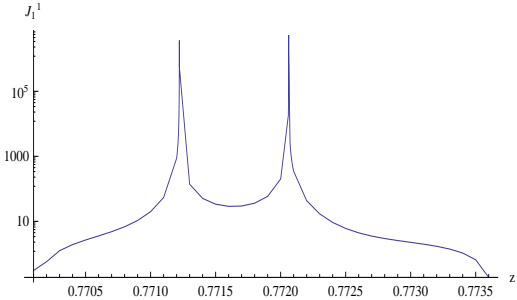}
\caption{A plot of $J^1 _{~1}$ in $eV^4$, versus $z$ (Case-IV) shows a couple of peaks at $z \approx 0.7712$ and $z \approx 0.7721$ where, $J^1_{~1} = 5.36 \times 10^6 ~eV^4$, i.e. $(J^1_{~1}) ^{\frac{1}{4}} \simeq T \approx 48 ~eV$, which is sufficient to reionize intergalactic medium.}\label{figure 6}
\end{center}
\end{figure}

\subsection{Does late time radiation era reionizes IGM?}
Present discrepancy between the abundance of galactic subhaloes predicted by N-Body simulations with those observed in the `Local Group', suggests an early reionization on the baryonic components of the universe. If reionization is described as an instantaneous increment of the intergalactic medium (IGM) temperature, a key role is supposed to have been played by Compton cooling at redshift $z > 10$, which counteracts any heating of the gas. A late reionization is therefore required at $z_{reion} < 10$ to sufficiently reduce the number of luminous dwarf satellites around our Galaxy. The temperature at this epoch increases up to $31.6~eV$, which is sufficient to ionize hydrogen. The absorpion spectra of SDSS (Sloan Digital Sky Survey) quasars at $z \sim 6$ indicate that the neutral fraction of hydrogen (reionization energy being $13.6$ eV) increases significantly at $z > 6$ \cite{reion:1a}-\cite{reion:1f} and the UV spectrum of quasars implies that helium (reionization energy being $54.4$ eV) is fully ionized only recently, at $z \approx 2.7$. Since low energy CMBR photons ($T \sim 0.4$ eV at $z \sim 1080$) falls far below reionization energy, it is usually assumed that the ultraviolet radiation and mechanical energy that preheated and reionized most of the hydrogen and helium in the IGM, ending the ``dark ages'', are due to early generation $(10 < z < 15)$ of subgalactic stellar systems (stars and Quasars) \cite{reion:2} aided by a population of accreting black holes \cite{mandau}. However, no equivocal resolution has yet been reached in this regard and the issue is still a mystery \cite{reion:3a}-\cite{reion:3e}. \\

\noindent
In subsection 7.1, we have shown that the modified theory of gravity under consideration, may be looked upon as to induce an effective electro-magnetic field tensor $\mathbb{T_{\mu\nu}}$ in the field equation. This means other than scalar-tensor equivalence, higher order theory may also be looked upon as tensor-tensor equivalence and the photons corresponding to such a theory might well interact with the atoms to ionize them. The source of the gravitational wave calculated from the transverse-traceless part of the space-space component of the energy-momentum tensor has been found to produce strong enough gravitational waves ($\sim 60~$eV, for $\beta > 0$, figure-5) to ionize both hydrogen and helium (case-II) and ($\sim 48~$eV, for $\beta < 0$, figure-6) to ionize hydrogen (case-IV). Therefore there is an indication that the effective electro-magnetic field might be responsible for reionizing IGM. So, at least from classical point of view, modified theory of gravity shades some light in the issue of reionization of the IGM. Indeed, for a definite claim in this respect, it is required to specify a mechanism for producing the first sources or energy injection into the standard model plasma in the form of the actual production of UV photons from the gravitational sector. For this purpose, it is required to solve the Schr\"odinger equation in the presence of gravitational Waves, which is beyond the scope of the present work.

\section{Conclusion:}
$f(R)$ theory of gravity has been taken seriously in recent years, to explain late time cosmological evolution. Here an action containing a linear term a curvature squared term and a three-half term together with baryonic matter has been taken into account to describe the cosmological evolution. Starobinsky inflation is one of the minimal models which naturally explains inflation and reheating from geometry itself, without invoking a scalar field. In fact, Starbinsky inflation is so powerful that adding three Majorana fermions to the standard model, it is possible to explain neutrino oscillation, inflation, reheating, dark matter generation and baryon asymmetry of the universe \cite{panin}. The importance of $R^{\frac{3}{2}}$ term has already been established, since no other form of $f(R)$ is admissible in view of Noether symmetry. When such a term $(R^{3\over 2})$ is added to the Starobinsky term, rest of the history after initial stage of the cosmological evolution is explained naturally. A smooth and continuous transition from early radiation era via matter-radiation equality ($z \approx 3200$), decoupling ($z \approx 1100$) through to late time cosmic acceleration after a long matter dominated era is clearly visible from the graphs. Additionally, a late time radiation era has also been observed, which is of particular interest. Since field equation (\ref{19}) shows that a part of $R^{\frac{3}{2}}$ term clearly acts like an effective energy-momentum tensor of an electromagnetic field, therefore this late time radiation era might be responsible for reionizing IGM. However, we have not presented a solid proof in this connection by solving Schr\"odinger equation in the presence of gravitational wave equation. Nevertheless, if gravitational wave corresponding to modified theory of gravity is responsible for reionization then, in the present model it ends at around $z \approx 3$ for $\beta > 0$ and $z \approx 0.8$ for $\beta < 0$, keeping all other cosmological data at par with observations.\\

\noindent
The seven year CMBR data has presented reionization result whose profile is a smooth ramp in the redshift space and the parameter $\Delta_z$ changes the slope of the ramp about its midpoint in such a way as to preserve total optical depth. Adding $\Delta_z$ as a parameter to the basic $\Lambda$CDM model and varying it in the range $0.5 < \Delta_z < 15$, the redshift of reionization has been found to be $z_{reion} = 10.5 \pm 1.2$ \cite{larson}, using CAMB (Code for Anisotropies in the Microwave Background) \cite{lewis}. Our result fits perfectly with such data as all the graphs show reionization epoch starting at around $z_{reion} \approx 10$, ($q > 0.5$). The peak depicts the end of reionization at $z \approx 3$ for $\beta > 0$ and $z \approx 0.8$ for $\beta < 0$, which fit earlier data \cite{reion:3a}-\cite{reion:3e}. The source of the gravitational wave calculated from the transverse-traceless part of the space-space component of the energy-momentum tensor shows that its strength is sufficient to reionize both hydrogen and helium in the IGM.\\

\noindent
Weak energy limit of the model under consideration has also been established following canonical formulation of the model with tensor-tensor mode and scalar-tensor mode. The mass of the scalaron in the second case shows Newtonian correction is insignificant in the solar system, since the Compton wavelength is very small (of the order of nanometre) while it falls within the limit of quantum stability bound in the bulk.\\

In view of all these, we conclude that Modified theory of gravity should be taken up even more seriously to understand if the primary investigations done here are relevant.\\

\section{Appendix(Calculation for weak energy limit):}
Field equation for action (4) reads,
 \begin{equation}\begin{split}\label{26}&
 R^{\frac{3}{2}} \left(R_{\mu \nu }-\frac{1}{2}R g_{\mu \nu }\right)+\beta_1 R^2 \left(3 R_{\mu \nu }-R g_{\mu \nu }\right)\\&
  +\frac{3}{2} \beta_1 g_{\mu \nu}\left(R R^{;\lambda}_{~;\lambda}-\frac{1}{2} R_{;\lambda}R^{;\lambda}\right)\\&
  -\frac{3}{2} \beta_1 \left(R R_{;\mu ;\nu }-\frac{1}{2} R_{;\mu}R_{;\nu }\right)=R^{\frac{3}{2}}T_{\mu \nu }
\end{split}\end{equation}
The trace of the above equation being multiplied by $\frac{1}{2}g_{\mu \nu}$, leads to
\begin{equation}\begin{split}\label{27}&
-\frac{1}{2}g_{\mu \nu}R^{\frac{5}{2}}-\frac{1}{2}g_{\mu \nu}\beta_1 R^3+\frac{9}{4}\beta_1 g_{\mu \nu}R \Box R\\&
-\frac{9}{8}\beta_1 g_{\mu \nu}R_{;\sigma}R^{;\sigma}=\frac{1}{2}R^{\frac{3}{2}}T g_{\mu \nu}
\end{split}\end{equation}
Now, the (0,0) component of the difference of equations (\ref{26}) and (\ref{27}) when divided by $R^{\frac{3}{2}}$, finally gives
\begin{equation}\label{28}\begin{split}&
 R_{0 0 }+\beta_1 R^{\frac{1}{2}} \left(3 R_{0 0 }-\frac{R}{2} g_{0 0 }\right)-\frac{3}{4} \beta_1 g_{0 0}R^{-\frac{3}{2}}\left(R \Box R\right.\\&
 \left.+R_{;\sigma}R^{;\sigma}\right) -\frac{3}{2} \beta_1 R^{-\frac{3}{2}} \left(R R_{;0 ;0 }-\frac{1}{2} R_{;0}R_{;0 }\right.\\&
 \left.-\frac{3}{4}g_{0 0}R_{;\sigma}R^{;\sigma}\right)=T_{0 0 }-\frac{1}{2}T g_{0 0}\end{split}
\end{equation}
Assuming only linear term, one can use the following relations in equation ({\ref 28})
\[R_{00}=\frac{1}{2}\nabla^2 h_{00},R=\frac{1}{2}\nabla^2 h,R_{;0}R^{;0}=0,g_{00}=(1+h_{00})\]
 \[\Box R=\frac{1}{2}
 \left(\nabla^2 h_{,i,j}\eta^{ij} +h^{ij}{_{,j}}\nabla^2 h_{,i}+\nabla^2 h_{,i}\eta^{ij}(\ln\sqrt{-g})_{,j}\right)\]
 \[R_{,\sigma}R^{,\sigma}=\frac{1}{4}\eta^{ik}\nabla^2 h_{,i}\nabla^2 h_{,k},~~~~~  R_{;0;0}=\frac{1}{4} \eta^{ik} h_{00,k}\nabla^2 h_{,i}
 \]
 to obtain,
\be
\begin{split}&
 \frac{1}{2}\nabla^2 h_{00}+\beta_1 (\nabla^2 h)^{\frac{1}{2}} \left(\frac{3}{2} \nabla^2 h_{0 0 }-\frac{1}{4}\nabla^2 h (1+h_{0 0 })\right)\\&
 -\frac{3}{4} \beta_1 (1+h_{0 0})(\frac{1}{2}\nabla^2 h)^{-\frac{3}{2}}\left(\frac{1}{4}\nabla^2 h\left(\nabla^2 h_{,i,j}\eta^{ij}+\right.\right.\\&
 \left.\left.h^{ij}{_{,j}}\nabla^2h_{,i}+\nabla^2h_{,i}\eta^{ij}(ln\sqrt{-g})_{,j}\right) +\frac{1}{4}\eta^{ik}\nabla^2 h_{,i}\nabla^2 h_{,k}\right)\\&
 -\frac{3}{2} \beta_1 (\nabla^2 h)^{-\frac{3}{2}} \left(\frac{1}{8} \eta^{ik} h_{00,k}\nabla^2 h\nabla^2 h_{,i}-0\right.\\&
 \left.-\frac{3}{16}\eta^{ik}\nabla^2 h_{,i}\nabla^2 h_{,k}\right)=T_{0 0}\end{split}
\ee
Now considering only linear term and next higher order terms, one finally obtains
\begin{equation}\label{29}
\triangledown^2 h_{00} + 3\beta_1 ~\sqrt{\frac{1}{2}\triangledown^2 h}~\Big(\triangledown^2 h_{00}-\frac{1}{6}\triangledown^2 h\Big) \simeq \rho.
\end{equation}
On the other hand, if we consider only linear term, Poisson equation is obtained
\begin{equation}\label{30}
\triangledown^2 h_{00}\simeq \rho.
\end{equation}

\noindent B. Modak and Kaushik Sarkar acknowledge PURSE, DST (India) and RGNF, UGC (India) respectively for financial support.\\

\end{document}